\documentstyle[prb,twocolumn,aps,epsf,floats]{revtex}

\newcommand{\co}{$\rm CuO_2$}

\newcommand{\be}{\begin{eqnarray}}
\newcommand{\ee}{\end{eqnarray}}

\begin{document}
%
\title{Thermodynamics of the low-temperature structural transition in
      rare-earth-doped La$_{2-x}$Sr$_x$CuO$_4$} 

\author{Ralph Werner\thanks{Present address: Physics Department,
Brookhaven National Laboratory, Upton, NY 11973-5000, USA.}}
\address{Institut f\"ur Physik, Universit\"at Dortmund,
         D-44221 Dortmund, Germany}
\author{M. H\"ucker and B. B\"uchner}
\address{II. Physikalisches Institut, Universit\"at zu K\"oln,
         D-50937 K\"oln, Germany}

\date{\today}
\maketitle

\begin{abstract}
The Sr concentration dependence of the structural transition from
the orthorhombic into the tetragonal low-temperature phase in
rare-earth-doped ($R$ = Nd, Eu) La$_{2-x-y}$Sr$_x$$R_y$CuO$_4$ has been
studied in detail. Independently of the rare earth concentration the
transition temperature is strongly reduced at $x\sim 0.05$. We give
a qualitative argument that the effect can be attributed to the
Coulomb repulsion between doped carriers. We find the enthalpy jump at
the low-temperature transition to scale transition temperature between
the high-temperature tetragonal and the orthorhombic phase. This
effect can be understood in terms of a Landau expansion.
\end{abstract}



\section{Introduction}

The observation of static stripe correlations in 
$\rm La_{1.475}Nd_{0.40}Sr_{0.12}CuO_4$ renewed the interest in
rare-earth-doped ($R$) lanthanide copper oxides.\cite{Tranquada95Nd} 
The observation of static stripes is induced by a low-temperature (LT)
structural transition from a low-temperature orthorhombic (LTO) to a
tetragonal low-temperature phase (LTT). A static superlattice
structure appears close to the structural transition, which can be
accounted for by the appearance of evenly spaced static
stripes.\cite{Zimmermann98,Niemoeller99,Baberski98} On the other hand,
the incommensurability of the magnetic 2D correlations seen in neutron
scattering experiments\cite{Mason92,Shirane94} with
La$_{2-x}$Sr$_{x}$CuO$_4$ can be viewed as a consequence of the 
formation of charged stripes separating antiferromagnetically ordered
domains, because the magnetic order parameter suffers an antiphase
jump across such a domain wall.\cite{Schulz89wand,Zaanen89wand} The
Sr dependence of the antiferromagnetic incommensurability is
consistent with the gathering of the doped carriers in domain
walls.\cite{Yamada97} In La$_{2-x}$Sr$_{x}$CuO$_4$ the domain walls
are expected to be very mobile objects,\cite{Zaanen96} rendering the
experimental proof of their existence extremely difficult. 

The antiferromagnetic incommensurability is assumed to be the same in
the LTO and LTT phase.\cite{Yamada97} Since in the LTT phase the
incommensurability can be accounted for by the static stripes the
obvious conclusion is that in the LTO phase the incommensurability
may be caused by dynamical stripes.\cite{Zaanen96} A better
understanding of the influence of the LTO-LTT transition on the
electronic system is thus desirable.   

The LTO-LTT transition is invoked by further rare-earth ($R$) doping
of La$_{2-x-y}$Sr$_{x}R_y$CuO$_4$.\cite{Axe94} 
In both phases the $\rm CuO_6$ octahedra are tilted with respect to
the crystallographic axes by an angle $\Phi<5^\circ$. In the LTO phase
the tilt axis is parallel [110]$_{\rm HTT}$ and rotates
discontinuously by $\theta=45^\circ$ towards the [100]$_{\rm HTT}$
direction at the transition into the LTT phase (we use the notation of
the undistorted tetragonal high-temperature phase HTT). Note that the
tilt angle $\Phi$ is roughly the same in the LTO and the LTT phase
while the different directions of the tilt produce different buckling
pattern of the $\rm CuO_2$ plane. The transition temperature is of the
order of $T_{\rm LT}\sim 100$ K. 

\begin{figure}[tb]
\epsfxsize=0.5\textwidth
\centerline{\epsffile{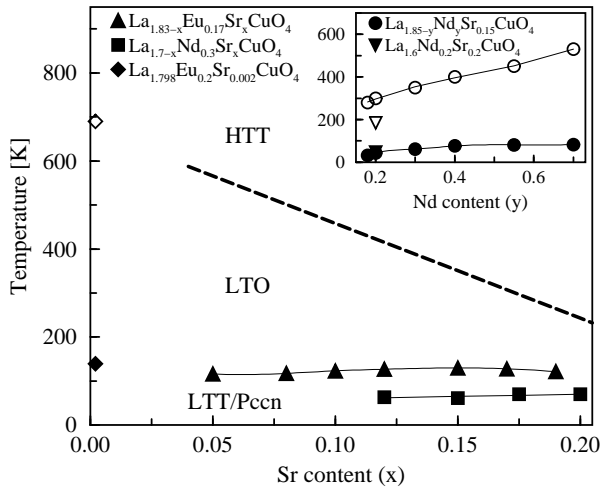}}
\centerline{\parbox{\textwidth}{\caption{\label{PhasesTO} \sl
Generic phase diagram obtained from the doping dependent transition
temperatures for different $R$ concentrations. Open symbols mark
HTT-LTO transition temperatures $T_{\rm HT}$, the broken line marks
the HTT-LTO transition for the $\rm Eu_{0.17}$ ($\circ$) and the $\rm
Nd_{0.3}$ ($\Box$) samples determined in earlier
work. (The stichometry has been chosen
such that both the $\rm Eu_{0.17}$ and $\rm Nd_{0.3}$ samples have the
same $T_{\rm HT}$.) The filled symbols show the LTO-LTT transitions
$T_{\rm LT}$. The figure shows the Sr dependence of the transitions,
the inset the Nd dependence. Thin lines are guides to the eye.}}} 
\end{figure}

In this work we investigate both Eu and Nd doped
La$_{2-x}$Sr$_x$CuO$_4$. It is worthwhile to mention that the LT phase
is the more stable the smaller the $R$ element, which renders the
europium samples advantageous. Especially for Nd doped
La$_{2-x}$Sr$_x$CuO$_4$ for $x<0.10$ the LTO-LTT transition becomes a
sequence of two transformations.\cite{Axe94} First at $T_{\rm LT}$ a
discontinuous transition from the LTO into the intermediate $Pccn$
phase takes place followed by a continuous development into the LTT
phase. The $Pccn$ phase is characterized by a  rotation of the tilt
axis which is less than $45^\circ$. In the Eu-doped samples the LT
transition occurs at higher temperatures ($T_{\rm LT}\sim 130$ K) and
the appearance of the $Pccn$ phase is strongly reduced to Sr contents
smaller $x\sim 0.07$. Yet, at lower temperatures ($\sim$50 K-60 K) all
samples are in the LTT phase making a direct comparison possible. The
generic phase diagram is shown in Fig.~\ref{PhasesTO}. 

The structural data have been collected on polycrystalline material
of Nd and Eu doped La$_{2-x}$Sr$_x$CuO$_4$. The samples used in this
study have been prepared by a standard solid state reaction described
by Breuer {\it et al.}\cite{Breuer98} Those with small Sr content
where annealed in $N_2$ atmosphere at temperatures between 550$^\circ$
and 625$^\circ$ for several days to remove excess oxygen.


\section{Characteristic minimum of the doping dependent
LTO-LTT transition} 

\begin{figure}[tb]
\epsfxsize=0.48\textwidth
\centerline{\epsffile{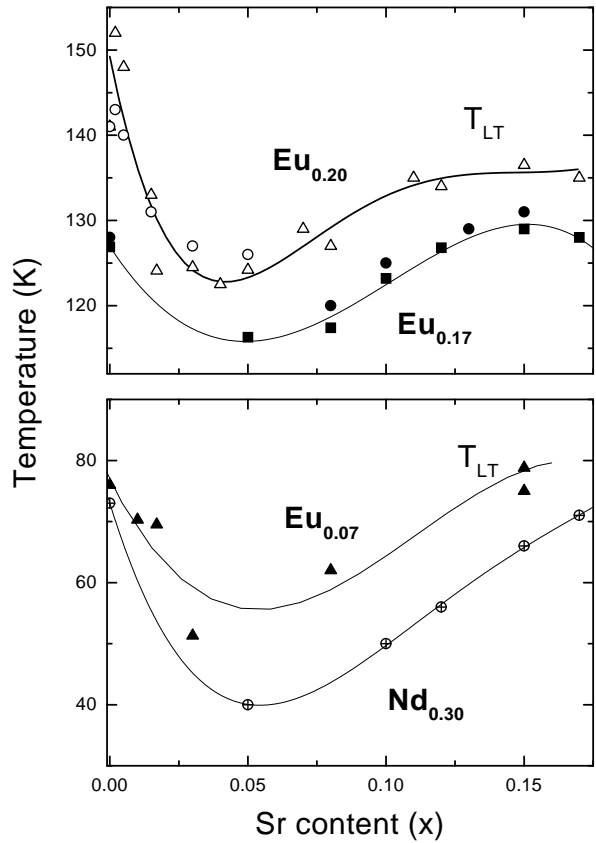}}
\centerline{\parbox{\textwidth}{\caption{\label{TLTofx} \sl LTO-LTT
transition temperature $\rm T_{\rm LT}$ in
La$_{2-x-y}$Sr$_xR_y$CuO$_4$ versus Sr content for four series of
samples as symbolized by: 
                $\bf Eu_{0.20}$: $\circ$ x-ray, $\triangle$ $\chi$; 
                $\bf Eu_{0.17}$: $\bullet$ x-ray, $\Box$ c$_p$;
                $\bf Eu_{0.07}$: $\triangle$ $\chi$;
                $\bf Nd_{0.30}$: $\oplus$ x-ray. 
                (Eu LTO-LTT and LTO-$Pccn$); (Nd LTO-$Pccn$)
}}}
\end{figure}

In Fig.~\ref{TLTofx} we present the Sr concentration dependence of
$T_{\rm LT}$ obtained from four series of samples with different Nd
and Eu concentrations. The data clearly shows that $T_{\rm LT}$ is a
nonmonotonous function of the hole content $x$. $T_{\rm LT}$ varies
between 70 K ($x=0$) and 153 K ($x=0.002$) for the samples with
extremely small Sr content, then strongly decreases by about 20 K to
30 K showing a local minimum around $x\sim 0.05$, and increases for
higher Sr concentrations. The transition temperatures reach a plateau
at intermediate doping $x>1/8$ and fall off again when approaching
the HTT-LTO transition line shown in Fig.~\ref{PhasesTO}. Note that
this decrease is not for all samples within the plot range of
Fig.~\ref{TLTofx}.

The minimum in the underdoped region centered around $x\sim 0.05$ is
characteristic for all $R$ doping. It is straightforward to attribute
this universal behavior to the same origin. The Sr doping strongly
effects the electron configuration of the CuO$_4$ planes. Also, an
explanation of the nonmonotonous behavior by sterical reasons fails
because no comparable Sr dependence is observed for other relevant
lattice parameters.\cite{Buechner93LTT,Buechner94LTT} Therefore, a
correlation with electronic properties is very
likely.\cite{Baberski98} 

\begin{figure}[tb]
\epsfxsize=0.5\textwidth
\centerline{\epsffile{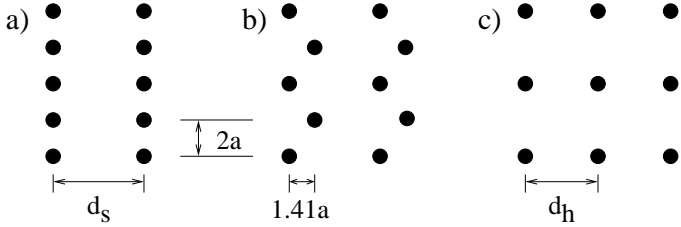}}
\centerline{\parbox{\textwidth}{\caption{\label{config} \sl
Hole configurations considered. a) Half filled static stripes in the
LTT phase. b) Zig-zag modulation of dynamic stripes with larger mean
distances between holes in the LTO phase. c) Homogeneous LTO
configuration. The distances $d_h=\sqrt{x}^{-1}$ and $d_s=(2x)^{-1}$
for the Sr concentration $x$ are given in units of the lattice spacing
$a$.}}} 
\end{figure}

We limit ourselves to present a qualitative argument showing that the
Coulomb repulsion between doped carriers is a possible origin of the
characteristic minimum of $T_{\rm LT}$. In the LTT phase we may
assume the holes to form static stripes. From the position of the
incommensurate magnetic and charge peaks of the neutron scattering
data of the La$_{1.48}$Nd$_{0.4}$Sr$_{0.12}$CuO$_4$ compound one can
conclude that the stripes carry one hole per two copper
sites.\cite{Tranquada95Nd} From the incommensurability measured for
other Sr concentrations\cite{Yamada97} we can assume that the half
filled case (1/2 hole per copper) is generic for pinned charged
stripes. The resulting static hole configuration is shown in
Fig.~\ref{config} a). 

In the LTO phase the hole configuration is still controversial. The
picture of a uniform charge distribution can be modeled by a static
square lattice formed by the holes as shown in Fig.~\ref{config}
c). Fluctuating stripes are represented by a zig-zag configuration of
the holes as shown in Fig.~\ref{config} b) accounting for the larger
mean distance between holes due to fluctuations. Configuration b) may
also be realized in the LTT phase by allowing fluctuations around the
static position of the stripe. 

We now compare the corresponding Madelung energies in the LTO and LTT
phase. 
\begin{equation}\label{Madelung}
\Delta E_c=x\sum_{{\bf r}_{\rm LTO}}
\frac{e^{-\lambda|{\bf r}_0-{\bf r}_{\rm LTO}|}}
                                      {|{\bf r}_0-{\bf r}_{\rm LTO}|}
          -x\sum_{{\bf r}_{\rm LTT}}
\frac{e^{-\lambda|{\bf r}_0-{\bf r}_{\rm LTT}|}}
                                      {|{\bf r}_0-{\bf r}_{\rm LTT}|}
\end{equation}
Here ${\bf r}_{\rm LTO}$ has to be summed over the sites occupied by the
holes in the homogeneous phase or on the zig-zag stripes and ${\bf
r}_{\rm LTT}$ over those in the static stripes. The factor $x$ is the hole
concentration and $\Delta E_c$ is thus the energy difference per Cu site. 
We have introduced the inverse Thomas Fermi screening length
$\lambda\sim 1/(2a)$ of the order of the lattice spacing $a$.
The sums are calculated using the mean distances of the holes
$d_h=\sqrt{x}^{-1}$ in the homogeneous case and the mean distance
between the stripes $d_s=(2x)^{-1}$ in units of the lattice spacing
$a$ (see Fig.~\ref{config}). The distance within the static stripes is
two lattice spacings.    

\begin{figure}[tb]
\epsfxsize=0.5\textwidth
\centerline{\epsffile{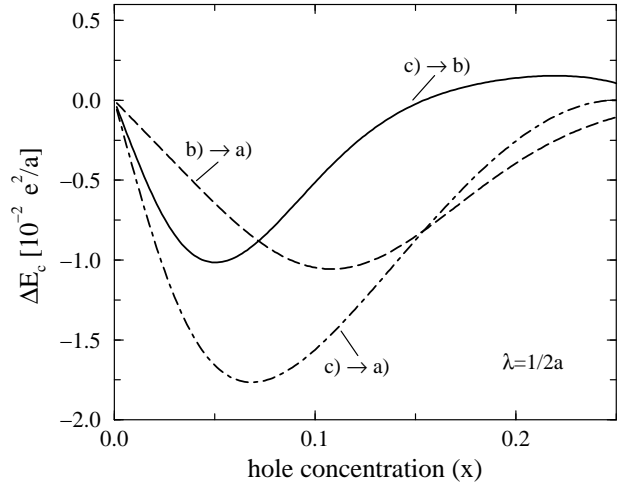}}
\centerline{\parbox{\textwidth}{\caption{\label{DeltaEc} \sl
Doping dependent differences of the Madelung energy per Cu site
between the striped LTT phase and the two LTO configurations shown in
Fig.~\protect\ref{config}. They are calculated according to
Eq.~(\ref{Madelung}). The energy is given in units of
$10^{-2}e^2/a\approx 4\cdot 10^{-2}$ eV. $a$ is the lattice
spacing. The positions of the minima are in qualitative agreement with
the experimental LTO-LTT transition lines
(Fig.~\protect\ref{TLTofx}).}}}    
\end{figure}

The resulting energy differences are shown in Fig.~\ref{DeltaEc}. They
vanish for small doping concentrations as well as at quarter
filling. Then the stripes are so close that the configurations are
close to equivalent. For intermediate doping the energy loss due to
the charge condensation in the stripes gives minima at $0.05<x<0.1$ in
agreement with the experimental situation. The suppression of the 
LTO-LTT transition can thus be explained by the Coulomb repulsion of
the holes. This is consistent with the formation of the intermediate
$Pccn$ phase in this doping regime. 

The comparison of the position of the minima with the experimental
data shows the best agreement with the picture of the formation of
fluctuating stripes out of a uniform phase (solid line in
Fig.~\ref{DeltaEc}). Notice though that for a lattice spacing of
$a\approx 3.8$ {\AA} the energy scale is given by $10^{-2}e^2/a\approx
4 \times 10^{-2}$ eV. This is much too large in comparison with the
enthalpy liberated across the transition, which is of the order of
$10^{-4}$ eV per Cu site (see Fig.~\ref{DHexp}). The actual physical
situation must thus be more involved.

\section{Enthalpy scaling}

\begin{figure}[tb]
\epsfxsize=0.5\textwidth
\centerline{\epsffile{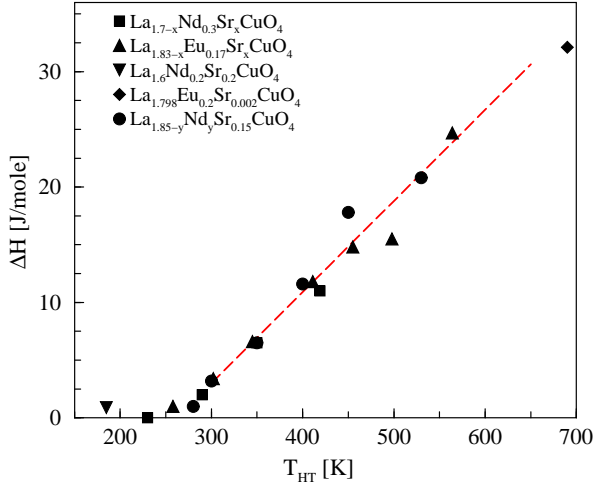}}
\centerline{\parbox{\textwidth}{\caption{\label{DHexp} \sl
Discontinuity of the enthalpy across the LT transition as a function
of the corresponding $T_{\rm HT}$. Full symbols are experimental values
from specific heat measurements. The broken line is a guide to the
eye.}}}
\end{figure}

For the samples shown in the phase diagram Fig.~\ref{PhasesTO}
specific heat measurements have been made. The area of
the anomaly at the transition determines the entropy $\Delta S$
liberated at the transition.\cite{*} In Fig.~\ref{DHexp} we have
plotted the resulting enthalpy $\Delta H= T_{\rm LT} \Delta S$ (filled
symbols) as a function of the transition temperature $T_{\rm HT}$ of the
HTT-LTO transition as shown in Fig.~\ref{PhasesTO}. The points all
fall roughly of the same line, even though we analyzed different
dopants varying both the Sr content $x$ and the Nd or Eu concentration
$y$. This implies a general relation between the HT and the LT
transition and a common driving mechanism for the structural
transitions. 

Note that within the accuracy of the experiment the electron
concentration in the CuO$_2$ planes does not play a crucial role for
the value of $\Delta H$. The La$_{1.85-y}$Nd$_y$Sr$_{0.15}$CuO$_4$
sample with constant hole doping ($\bullet$ in Fig.~\ref{DHexp})
follows the common line. This is consistent with the fact that
structural phase transitions involving an octahedron tilt are common
amongst perovskites.\cite{BruceCowley81}

\subsection{Landau theory}

To obtain a phenomenological theoretical understanding we apply the
standard Landau expansion of Gibb's free enthalpy in two order
parameters:\cite{Axe89} 
\begin{equation}\label{LandauQ1Q2}
G=a(Q_1^2+Q_2^2)+2b(Q_1^4+Q_2^4)+ 4c\, Q_1^2Q_2^2\,.
\end{equation}
The order parameters $Q_1$ and $Q_2$ measure the tilt of the $\rm
CuO_6$ octahedra about the [110]$_{\rm HTT}$ and
[1$\overline{1}$0]$_{\rm HTT}$ axis, respectively. The tilt axes are
symmetric (lattice invariant under rotation of $90^\circ$), thus the
coefficients are identical for both parameters. Since the tilt angle
$\Phi < 5^\circ$ is small, we disregard higher then quartic
terms. Applying the transformation $Q_1=Q\cos\theta$ and
$Q_2=Q\sin\theta$ we obtain new order parameters $Q$ measuring the
size of the octahedron tilt angle $\Phi$ and the angle $\theta$
measuring the direction of the tilt with respect to the [110]$_{\rm
HTT}$ axis. 
\begin{equation}\label{LandauQtheta}
G=a\, Q^2+ b\, Q^4 + c\, Q^4 \sin^2(2\theta)
\end{equation}
Above $T_{\rm LT}$ $\theta$ vanishes and $G$ reduces to a single order
parameter expansion. The change of sign of the second order coefficient
induces the HTT-LTO transition with the transition temperature given
by $a(T_{\rm HT})=0$. 

The temperature dependence of the order parameter in the LTO phase is
usually expressed via 
\begin{equation}\label{QofT}
Q(T)=Q_0(T_{\rm HT}-T)^\nu\,.
\end{equation}
In a mean field picture we have $a(T)=a_0(T_{\rm HT}-T)$, $b(T)=$ const,
and $\nu=0.5$. The LTO-LTT transition is of first order induced by a
jump of $\theta$ to a finite value of $45^\circ$. The transition
temperature is determined by $c(T_{\rm LT})=0$, i.e., the coefficient
changes sign. For small Sr concentrations $x<0.1$ in some samples the
jump in $\theta$ is smaller and $45^\circ$ are only reached after
further cooling, giving rise to an intermediate $Pccn$ phase. Within
this approach we neglect this effect. This is justified since the
enthalpy discontinuity show in Fig.\ \ref{DHexp} was obtained by
integrating over the whole specific heat anomaly down to the LTT
regime. 

The potential $G$ is continuous across the transition but its
derivative with respect to temperature is not, which leads to a
discontinuity in the entropy. 
\begin{equation}\label{DeltaS}
\Delta S=\frac{\partial G}{\partial T}\Big|_{T=T_{\rm LT}^+}-
\frac{\partial G}{\partial T}\Big|_{T=T_{\rm LT}^-}=
          \left[ Q^4\sin^2(2\theta)\frac{\partial c}{\partial T}
          \right]_{T=T_{\rm LT}^-}
\end{equation}
For this result we have assumed the derivatives 
$(\partial a)/(\partial T)$, $(\partial b)/(\partial T)$, 
and $(\partial Q)/(\partial T)$ to be continuous across the
transition, and we used the fact that $c(T=T_{\rm LT}^-)=0$ and
$\theta(T>T_{\rm LT})=0$. For the discontinuity of the enthalpy we obtain
\begin{equation}\label{DeltaH}
\Delta H=T_{\rm LT}\,\Delta S=
          T_{\rm LT}\ Q^4(T_{\rm LT})\ \frac{\partial c}{\partial T}
          \Big|_{T=T_{\rm LT}^-}\,.
\end{equation}

\subsection{Determination of $Q(T)$}

\begin{figure}[tb]
\epsfxsize=0.5\textwidth
\centerline{\epsffile{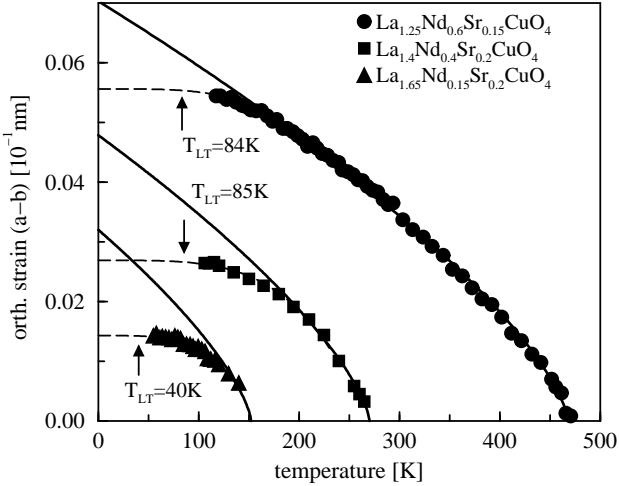}}
\centerline{\parbox{\textwidth}{\caption{\label{OP} \sl Orthorhombic
strain $(a-b) = Q^2 \sim \Phi^2$. $a$ and $b$ are the
lattice constants along the respective crystallographic axes determined
by x-ray scattering. The full lines are fits using Eq.\
(\protect\ref{QofT}) with $2\nu=0.7$ and $Q_0=0.031 \sqrt{{\rm
\AA}/{\rm K}^{0.7}}$ for all
samples. Broken lines are fits in the shape of $Q^2(T)=Q^2(0)\ (1-{\rm
const}\,T^4)$ to match the order parameter saturation below
$T^\ast\sim 150$ K to $200$ K.}}}    
\end{figure}

The temperature dependence of the octahedron tilt angle can be
specified experimentally via the orthorhombic strain $(a-b) = Q^2 \sim
\Phi^2$ as shown in Fig.\ \ref{OP}. Here $a$ and $b$ are the lattice
constants along the respective crystallographic axes determined by 
x-ray scattering. The full lines in Fig.\ \ref{OP} are given by the
square of Eq.\ (\ref{QofT}):
\begin{equation}\label{QofTfit}
Q^2(T)=0.00095\frac{{\rm \AA}}{{\rm K}^{0.7}}\ (T_{\rm HT}-T)^{0.7}.
\end{equation}
For all samples we set 
$Q_0=0.031 \sqrt{{\rm \AA}/{\rm K}^{0.7}}$. In agreement with
renormalization group studies\cite{Fisher74} an exponent of $2\nu=0.7$
satisfactorily reproduces the experimental values above the saturation
temperature $T^\ast\sim 150$ K to $200$ K. Note that obtaining the
universal prefactor $Q_0$ requires {\em not} to use reduced
temperatures $(T_{\rm HT}-T)/T_{\rm HT}$ in Eq.\
(\ref{QofTfit}).  

For all samples the LTO-LTT transition temperature is well below the
saturation temperature $T_{\rm LT}<T^\ast$. To quantify the saturation
behavior would require a systematic study of the saturation values of
$Q(T\to 0)$ which has not been done. At $T=T_{\rm LT}$ the deviation
from the fit given by Eq.\ (\ref{QofTfit}) is roughly of the same
magnitude for all samples and as a simple approximation we introduce a
constant correction term. 
\begin{equation}\label{QofTLT}
Q^2(T_{\rm LT})\approx 0.00095\frac{{\rm \AA}}{{\rm K}^{0.7}}\ (T_{\rm
                      HT}-T_{\rm LT})^{0.7} - 0.01\ {\rm \AA}
\end{equation}

\subsection{Comparison of theory and experiment}
 
\begin{figure}[tb]
\epsfxsize=0.5\textwidth
\centerline{\epsffile{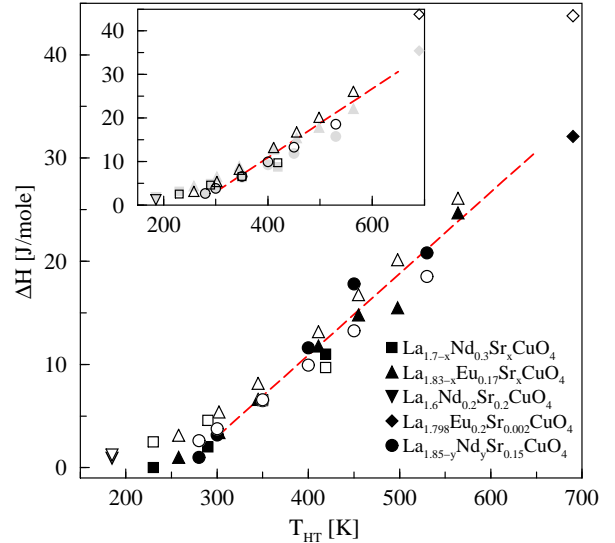}}
\centerline{\parbox{\textwidth}{\caption{\label{DHtheo} \sl
Theoretical discontinuity of the enthalpy across the LT
transition from Eq.\ (\protect\ref{DeltaH}) as a function of the
corresponding $T_{\rm HT}$. Open symbols in figure
and inset are identical and are estimated using Eq.\
(\protect\ref{QofTLT}) for the order parameter, grey full symbols in
the inset are obtained via Eq.\ (\protect\ref{QofTfit}). The black full
symbols and broken lines are identical to those in Fig.\
\protect\ref{DHexp} to allow for comparison.}}}      
\end{figure}

With the values of $\nu$ and $Q_0$ thus given and the data for $T_{\rm
LT}$ and $T_{\rm HT}$ presented in Fig.~\ref{PhasesTO} we calculate the
theoretical values for the enthalpy from Eq.~(\ref{DeltaH}). The open
symbols in Fig.~\ref{DHtheo} are the results calculating $Q(T_{\rm LT})$
via Eq.\ (\ref{QofTLT}). The scaling factor has been set to
$(\partial c)/(\partial T)|_{T=T_{\rm LT}^-}=66\ {\rm J}/({\rm mole\
K\ \AA}^2)$. The black full symbols and broken lines in Figs.\
\ref{DHexp} and \ref{DHtheo} are identical to allow a comparison
with the directly measured values for $\Delta H$. The inset in
Fig.~\ref{DHtheo} shows a comparison between values obtained obtained
by via Eq.\ (\ref{QofTLT}), open symbols, and those obtained without
correction for saturation from Eq.\ (\ref{QofTfit}), grey full
symbols. The scaling factor for gray symbols has been set to
$(\partial c)/(\partial T)|_{T=T_{\rm LT}^-}=41\ {\rm J}/({\rm mole\
K\ \AA}^2)$. In both cases the theoretical values show global
scaling. The open symbols satisfactorily fall on the the same line as
the directly measured data given in Fig.\ \ref{DHexp}. An exception is
the large value for La$_{1.798}$Eu$_{0.2}$Sr$_{0.002}$CuO$_4$ which
may be explained by an underestimated saturation of the order
parameter or by a deviation of the prefactor $Q_0$ due to excess
oxygen. We can thus conclude that the essential physics of the
structural phase transition is captured by the Landau approach. 

\subsection{Thermal conductivity and superconductivity}

The relation between lattice and electronic degrees of freedom has
already been addressed in the context of thermal
conductivity.\cite{Baberski98} As can be seen in the case of the
Nd$_{\rm 0.3}$ samples in Fig.\ \ref{DHexp} the enthalpy anomaly
becomes very small for a Sr content of $x\ge0.17$. For $x\ge0.17$ the
anomaly of the thermal conductivity $\Delta\kappa$ across the LT
transition also vanishes as shown in Ref.\
\onlinecite{Baberski98}. Comparing Eq.\ (\ref{QofTLT}) with the Sr
dependence of the anomaly of the thermal conductivity
$\Delta\kappa$,\cite{Baberski98} one can conclude that $\Delta\kappa$ 
is closely related to the octahedron tilt $\Delta\kappa\sim Q(T_{\rm
LT})$. The appearance of superconductivity in the LTT phase can
equally be attributed to a small enough octahedron tilt order
parameter\cite{Buechner94LTT} and occurs in the Nd$_{\rm 0.3}$ samples
also for $x\ge0.17$. Thus the octahedron tilt angle and phononic and
electronic transport properties are closely related.\cite{Baberski98}

In La$_{2-x}$Sr$_{x}$CuO$_4$ without additional rare-earth doping, the
$Z$-point phonon associated with the LTO-LTT transition softens only
down to the temperature where superconductivity
appears,\cite{Kimura00HTSC} underlining the electronic transport
dependence on the crystal dynamics. A possible relation between the
magnetic incommensurability and high-temperature lattice fluctuations
has also been pointed out.\cite{Kimura00HTSC}

\section{Conclusions}

We have presented experimental data showing a universal characteristic
minimum of the LTO-LTT transition temperature $T_{\rm LT}$ as a function
of the Sr content of the sample independent of the Eu or
Nd concentration. We have shown that the doping dependent maximum of
the loss of Coulomb energy through the formation of charged stripes is
a possible origin of the effect. 

Analyzing specific heat measurements we have shown that the enthalpy
jump at the  LTO-LTT transition shows universal scaling as a function
of the HTT-LTO transition temperature. Within the variation of the
results the scaling is independent of the electron concentration in
the \co\ planes, showing that the structural transitions are
essentially driven by the lattice dynamics. Results from a simple
Landau expansion are consistent with the scaling behavior. 

In conclusion our analysis of the LT transition does not unambiguously
show the relevance of electronic degrees of freedom for this phase
transition. Whereas the Sr-content dependence of $T_{\rm LT}$ suggests
the influence of electronic properties, our discussion of the enthalpy
reveals that the energy scale of the phase transition is basically
determined by sterical lattice properties.

\section{Acknowledgments}

R.\ Werner thanks C.\ Gros for discussions. The support of the
DFG is gratefully acknowledged.

\end{document}